\documentclass[aps,prb,twocolumn,floatfix,amsmath,amssymb]{revtex4}
\usepackage[final]{graphicx}
\usepackage{bm}
\usepackage{afterpage}
\usepackage{color}
\usepackage{comment}
\usepackage{bm}
\usepackage[colorlinks=true,urlcolor=blue,linkcolor=blue,citecolor=blue]{hyperref}

\emergencystretch=\maxdimen
\allowdisplaybreaks[4]

\begin{document}

\title{Hybrid higher-order skin-topological effect in hyperbolic lattices}

\author{Junsong Sun}
\affiliation{School of Physics, Beihang University,
Beijing, 100191, China}

\author{Chang-An Li}
\email{changan.li@uni-wuerzburg.de}
\affiliation{Institute for Theoretical Physics and Astrophysics, University of W$\ddot{u}$rzburg, 97074 W$\ddot{u}$rzburg, Germany}

\author{Shiping Feng}
\affiliation{ Department of Physics,  Beijing Normal University, Beijing, 100875, China}

\author{Huaiming Guo}
\email{hmguo@buaa.edu.cn}
\affiliation{ School of Physics, Beihang University,
Beijing, 100191, China}

\begin{abstract}
  We investigate the non-Hermitian Haldane model on hyperbolic $\{8, 3\}$ and $\{12, 3\}$ lattices, and showcase its intriguing topological properties in the simultaneous presence of non-Hermitian effect and hyperbolic geometry.  From bulk descriptions of the system, we calculate the real space non-Hermitian Chern numbers by generalizing the method from its Hermitian counterpart and present corresponding phase diagram of the model. For boundaries, we find that skin-topological modes appear in the range of the bulk energy gap under certain boundary conditions, which can be explained by an effective one-dimensional zigzag chain model mapped from hyperbolic lattice boundary. Remarkably, these skin-topological modes are localized at specific corners of the boundary, constituting a hybrid higher-order skin-topological effect on hyperbolic lattices.
\end{abstract}


\maketitle

\section{Introduction}
The non-Hermitian system is a non-conservative open system\cite{NHphysics}, in which many novel physical phenomena that have no counterpart in the Hermitian case can emerge. Complex eigenvalues are allowed in non-Hermitian systems, and  exceptional points at which both eigenvalues and eigenstates coalesce
can appear\cite{NHphysics,RevModPhys.93.015005}.  The coexistence of non-Hermitian effect and topology has resulted many exotic non-Hermitian phenomena being discovered and reinterpreted \cite{PhysRevX.9.041015,PhysRevLett.118.040401,PhysRevLett.125.118001,PhysRevB.97.121401,Loghi19prl,Budich20prl,Borgnia20prl,LiLH20prl,GuoC21prl,SunX21prl,PhysRevB.105.L100102,ZhangK22nc,PhysRevLett.122.076801,PhysRevLett.123.073601,PhysRevB.98.165148,PhysRevB.105.075128,PhysRevLett.124.236403}. Non-Hermitian skin effect\cite{PhysRevLett.121.086803,review_skin_effect,Okuma23review,PhysRevLett.125.186802} is a prominent example among them, that is, all eigenstates are localized to open boundaries of the non-Hermitian system. The non-Hermitian skin effect breaks down the usual bulk-boundary correspondence predicted by the Bloch theory \cite{PhysRevLett.121.086803,PhysRevLett.116.133903}. A non-Bloch theory  based on the generalized Brillouin zone\cite{PhysRevLett.121.086803,PhysRevLett.125.226402,PhysRevLett.121.136802,PhysRevLett.123.066404} is developed to remedy the bulk-boundary correspondence in non-Hermitian systems.  Interestingly, it is also found that the interplay between non-Hermitian skin effect and the topological boundary state can give rise to novel skin-topological modes\cite{PhysRevLett.123.016805}, so-called the hybrid higher-order skin-topological effect\cite{PhysRevLett.123.016805,PhysRevLett.128.223903,Kawabata20prb,Okugawa20prb,FuY21prb,ZhangX21nc,Okugawa21prb, ZouD21nc,ZhuW22prb,LiCA21arxiv}.

Recently, a new kind of lattice called hyperbolic lattice has attracted considerable interest\cite{PhysRevB.105.125118,RN55,RN56,RN57,Zhang23nc,PhysRevLett.125.053901,PhysRevLett.129.246402,PhysRevB.105.245301,2022arXiv220902904L,2022arXiv220902262T,doi:10.1073/pnas.2116869119,doi:10.1126/sciadv.abe9170,PhysRevLett.129.088002,Zhu_2021}. The hyperbolic lattice is a lattice that exists in a space of constant negative curvature. One can tessellate hyperbolic planes with any regular $p$-side polygon with $p>2$, unlike in a two-dimensional Euclidean plane where only three regular $p$-side ($p=3,4,6$) polygons tessellation can exist. With the experimental realization of hyperbolic lattices in circuit quantum electrodynamics\cite{RN55} and electric circuit\cite{RN56,RN57}, this greatly inspired the research interest in hyperbolic lattice, such as the topological properties of hyperbolic lattice\cite{PhysRevLett.125.053901,PhysRevLett.129.246402}, Chern insulator\cite{PhysRevB.105.245301}, high-order topological insulator\cite{2022arXiv220902904L,2022arXiv220902262T}, and hyperbolic band theory\cite{doi:10.1073/pnas.2116869119,doi:10.1126/sciadv.abe9170,PhysRevLett.129.088002} etc. So far, most of the research efforts on non-Hermitian effect focus on the lattice in Euclidean space but little attention has been paid on the corresponding effect in hyperbolic lattices. Stimulated by the experimental realization of hyperbolic lattices and rich topological physics in non-Hermitian systems, we explore the non-Hermitian topological effects on hyperbolic lattices in this work.

We extend the non-Hermitian Haldane model \cite{PhysRevLett.128.223903} to hyperbolic lattices (including hyperbolic $\{8,3\}$ and $\{12,3\}$ lattices), and systematically study the topological properties under non-Hermitian effects. By calculating the real-space non-Hermitian Chern number, we present a phase diagram for the model in related parameter space. We also verify the topological robustness of corresponding boundary states against Anderson disorder by the localization index called inverse participation ratio. As such,  we prove the exotic topological properties in the non-Hermitian Haldane model on both hyperbolic $\{8,3\}$ and $\{12,3\}$ lattices from bulk as well as boundary descriptions. Remarkably, we find that the non-Hermitian effect will drive the one-dimensional topological boundary states to  skin-topological modes at corners, forming a hybrid higher-order skin-topological effect. We explain the appearance of such a hybrid higher-order skin-topological effect by an one-dimensional zigzag chain model mapped from the specific boundary conditions of hyperbolic lattices.

The rest of this manuscript is organized as follows. Section~\ref{sec2} introduces the non-Hermitian Haldane model Hamiltonian. Section~\ref{sec3} investigates topological properties of non-Hermitian Haldane model and their robustness against disorder on hyperbolic $\{8,3\}$ lattice. Section~\ref{sec3} presents the appearance of hybrid higher-order skin-topological effect and its explanation. Section~\ref{sec4} generalizes our main results to hyperbolic $\{12,3\}$ lattice. Finally, we enclose some further discussion and conclusions in Sec.~\ref{sec5}.

\section{Non-Hermitian hyperbolic Haldane model}
\label{sec2}
We consider a non-Hermitian version of the celebrated Haldane model on hyperbolic lattice\cite{PhysRevLett.61.2015},
\begin{equation}\label{Eq1}
  \begin{aligned}
H=& t_1 \sum_{\langle i, j\rangle} c_i^{\dagger} c_j+t_2  \sum_{\langle\langle i, j\rangle\rangle}e^{i \nu_{i j} \phi} c_i^{\dagger} c_j \\
+&(m+{ i}\gamma)\sum_{i \in A} c_i^{\dagger} c_i-(m+{ i}\gamma)\sum_{i \in B} c_i^{\dagger} c_i,
\end{aligned}
\end{equation}
where $c_i^\dagger$ ($c_i$) is the creation (annihilation) operator of electron at site $i$. The first term is the nearest-neighbor(NN) hopping with the amplitude $t_1$. The second term represents a complex next-nearest neighbor (NNN) hopping with an amplitude $t_2$ and a phase $\nu_{ij}\phi$. The value of $\nu_{ij}$ is $+1(-1)$ for the hopping in the clockwise (counter-clockwise) direction. The phase $\phi$ is proportional to the flux enclosed by the cyclic NNN hoppings, and we set $\phi=\pi/2$ throughout the paper. $m$ and $i\gamma$ in the third term are real and imaginary on-site stagger potentials, respectively. The Hamiltonian is made non-Hermitian by the imaginary part of the potential, which is related to the on-site energy gain and loss.

\begin{figure}[htbp]
\centering
\includegraphics[width=8.7cm]{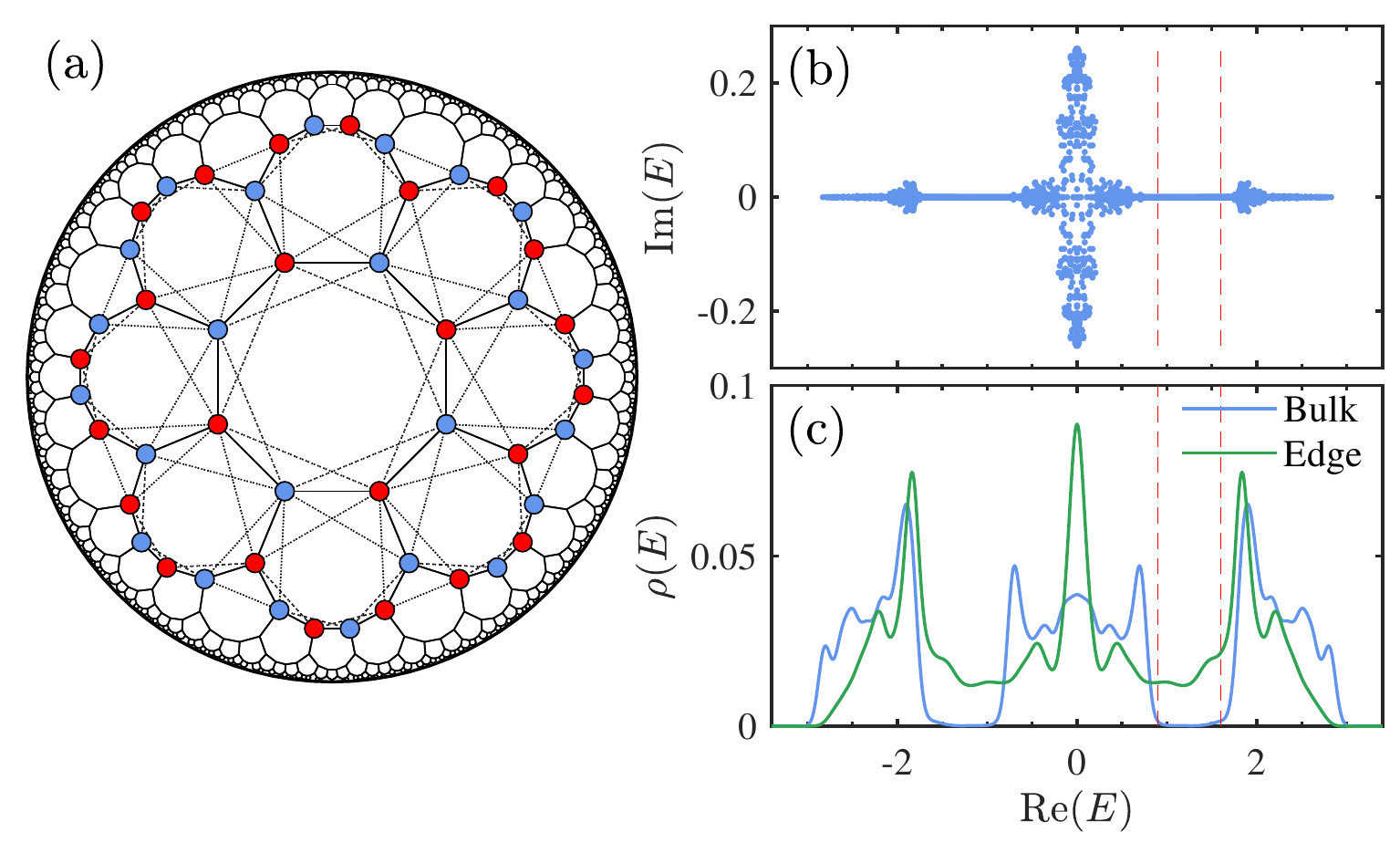}
\caption{(a) Schematic illustration of the Haldane model on a hyperbolic $\{8, 3\}$ lattice. The black dashed lines indicate the next-nearest neighbor hoppings with amplitudes $t_2e^{\pm i\phi}$. The sign of $\phi$ is positive (negative) for the (opposite) direction marked by the arrow, which arises due to an alternating magnetic flux. (b) Energy spectrum of the Hamiltonian in (1) on a hyperbolic $\{8, 3\}$ lattice with open boundary condition. (c) The corresponding density of states of bulk (blue) and edge (green) states. The red dotted lines in (b) and (c) estimate the topological gap above the energy $\mathrm{Re}(E)=0$ (the lower one is symmetric). Here the parameters in (b) and (c) are $t_1=1$, $t_2=0.2$, $m=0$ and $\gamma=0.3$.}\label{fig1}
\end{figure}
We first consider the Hamiltonian in Eq.~\ref{Eq1} on a hyperbolic $\{8, 3\}$ lattice [see Fig.~\ref{fig1}(a) for the geometry]. The Hermitian case, i.e., $\gamma=0$, has been examined in a recent work\cite{PhysRevLett.129.246402}. It is found that while the real staggered potential only results in a trivial insulator at half filling, the NNN spin-orbit coupling open two topological gaps located above and below $\mathrm{Re}(E)=0$, respectively. The resulting topological phases are characterized by non-trivial Chern numbers and chiral edge states.

Figure \ref{fig1}(b) plots the energy spectrum of a case of $\gamma \neq0$ and $m=0$ on a circular flake of hyperbolic $\{8,3\}$ lattice. Due to the non-Hermitian nature of the Hamiltonian, the eigenvalues are complex, thus are demonstrated as points in the $[{\rm{Re}}(E),{\rm{Imag}}(E)]$ plane. Under the basis $\Psi=(\psi_1^A,\ldots,\psi_N^A,\psi_1^B,\ldots,\psi_N^B)^\mathrm{T}$, the Hamiltonian in real space can be written as $H=\Psi^\dagger \mathcal{H}\Psi$. It can be verified that the matrix $\mathcal{H}$ satisfies a pseudo-Hermitian condition\cite{Mostafazadeh2001PseudoHermiticityVP},
\begin{equation}\label{Eq2}
  \eta \mathcal{H}\eta=\mathcal{H}^\dagger,
\end{equation}
where
\begin{equation}\label{Eq3}
  \eta=\left[
         \begin{array}{cc}
           0 & I_N \\
           I_N & 0 \\
         \end{array}
       \right],
\end{equation}
and $I_N$ is a $N\times N$ identity matrix. Suppose $\cal{H}$ has a complex eigenvalue $E_j$ with the eigenvector $\varphi_j$, which satisfy $E_j=\langle \varphi_j|{\cal{H}}|\varphi_j\rangle$. By a Hermitian operation, we get
\begin{equation}
E_j^*=\langle \varphi_j^*|{\cal{H}}^{\dagger}|\varphi_j^*\rangle=\langle \varphi_j^*|\eta{\cal{H}}\eta|\varphi_j^*\rangle.
\end{equation}
This implies $E_j^*$ is also an eigenvalue of $\cal{H}$ with the corresponding eigenvector $\eta|\varphi_j^*\rangle$.
Hence the complex eigenvalues appear in conjugate pairs, resulting the mirror-symmetric spectrum about the $x$-axis.

In the next, we calculate the corresponding density of states. Since a circular flake geometry is considered, we would like to distinguish the contributions from the bulk and edge states. For the $j$th eigenvector $|\varphi_j\rangle=\{ \phi_{j,x}\}_{x\in N_{\rm sites}}$, we define the bulk weight $\omega_j^{\rm bulk}$ and the edge one $\omega_j^{\rm edge}$:
\begin{align}
\omega_j^{\rm bulk}=\sum_{x\in N_{\rm bulk}}|\phi_{j,x}|^2,\quad\
\omega_j^{\rm edge}=\sum_{x\in N_{\rm edge}}|\phi_{j,x}|^2,
\end{align}
where $N_{\rm bulk} (N_{\rm edge})$ is the number of sites enclosed inside (outside) a circle of $0.95$ radius of the circular flake.
The bulk (edge) density of states can be extracted via
\begin{align}
\rho_{\text {b(e) }}(E)=\frac{1}{N_{\text {sites }}} \sum_{j=1}^{N_{\text {sites }}} \omega_j^{\text {bulk(edge) }} \delta_\eta\left(E-E_j\right),
\end{align}
where we choose a Gaussian smearing function to approximate the delta function $\delta_\eta(\varepsilon)=\frac{1}{\eta \sqrt{2 \pi}} \exp \left[-\frac{\varepsilon^2}{2 \eta^2}\right]$. As shown in Fig.~\ref{fig1}(c), the two symmetric energy gaps generated by the spin-orbit coupling remain even in the presence of a non-Hermitian term. Besides, there appear edge states traversing the gaps, implying the resulting non-Hermitian insulators are still topologically nontrivial.

The Hermitian hyperbolic topological band insulators have been characterized by the real-space Chern number formulated as
\begin{subequations}\label{Eq4}
\begin{align}
  C_{\mathrm{RS}}(\mu)&=12 \pi \mathrm{i} \sum_{j \in A} \sum_{k \in B} \sum_{l \in C}\left(\mathbb{P}_{j k}^\mu \mathbb{P}_{k l}^\mu \mathbb{P}_{l j}^\mu-\mathbb{P}_{j l}^\mu \mathbb{P}_{l k}^\mu \mathbb{P}_{k j}^\mu\right), \label{Za} \\
  \mathbb{P}^\mu&=\sum_{E_j<\mu} \left|\varphi^R_j\right\rangle\left\langle\varphi^L_j\right|, \label{Zb}
\end{align}
\end{subequations}
where $\mathbb{P}^\mu$ is the projector onto the subspace of occupied single-particle states at chemical potential $\mu$, and A, B, C are three regions in the bulk of the systems that do not extend all the way to the boundary, and which are arranged counter-clockwise around the center of the system. To apply the method to the non-Hermitian case, it is necessary to distinguish the left and right eigenvectors in constructing the projector $\mathbb{P}^\mu$, which satisfy $\mathcal{H}^\dagger\left|\varphi^L_j\right\rangle=E_j^*\left|\varphi^L_j\right\rangle$ and $\mathcal{H}\left|\varphi^R_j\right\rangle=E_j\left|\varphi^R_j\right\rangle$, respectively\cite{PhysRevLett.123.246801}.
Since the non-Hermitian Hamiltonian can be written as $\mathcal{H}=T\Lambda T^{-1}$ with $\Lambda$ a diagonal matrix, the right and left eigenstates are just the columns of $T$ and $(T^{-1})^\dagger$, and are orthonormal, i.e., $\left\langle\varphi^L_i|\varphi^R_j\right\rangle=\delta_{ij}$.

\begin{figure}[hbpt]
\centering
\includegraphics[width=8.6cm]{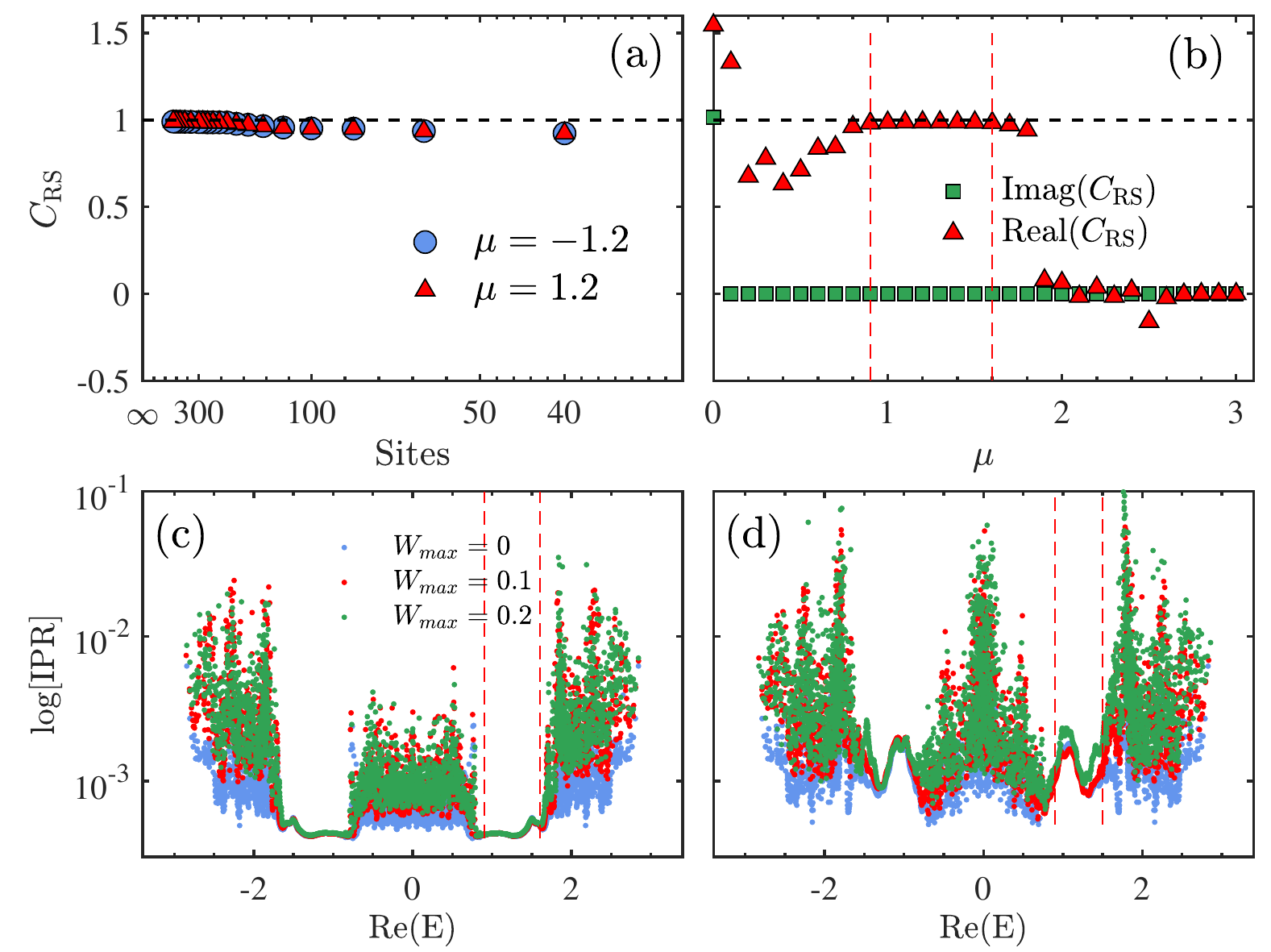}
\caption{(a) Convergence of the real-space Chern number for the two non-Hermitian topological phases as the lattice size increases. The blue (red) symbol represents the case that the chemical potential is located in the lower (upper) topological gap. (b) $C_{RS}$ as a function of chemical potential $\mu$ on a hyperbolic $\{8,3\}$ lattice with $500$ sites. IPR for various disorder strengths with (c) and without (d) the non-Hermitian potential. In (c) and (a,b,d), the values of the imaginary potential are $\gamma=0$ and $0.3$, respectively. The other parameters are the same as those in Fig. \ref{fig1}(b).}\label{fig2}
\end{figure}

Let us first calculate the topological invariant $C_{RS}$ when the chemical potential lies in the topological gap.
As shown in Fig.~\ref{fig2}(a), the Chern number converges to a quantized value as the size of the open hyperbolic lattice increases, indicating the non-Hermitian topological phase can be properly characterized by the definition in Eq.(7).
Then $C_{RS}$ as a function of chemical potential is obtained on a large enough lattice [see Fig.~\ref{fig2}(b)]. The real-space Chern number keeps the quantized value within the topological gap, but takes zero or a random value outside the bulk gap. Hence the appearance of the edge state is directly related to the quantization of the topological invariant, establishing the non-Hermitian bulk-boundary correspondence. It is noted the real-space Chern number varies continuously near the topological transition point, which should be due to the finite-size effect.

Generally, the edge states originating from the nontrivial topology are robust to the disorder that preserves the symmetry protecting the topological phase. To verify the robustness of the non-Hermitian topological edge states, we introduce a random on-site potential on each site
\begin{align}
H_{\rm dis}=\sum_{j=1}^{N_{\rm sites}}U_jc_i^{\dagger}c_i,
\end{align}
where the strength of the Anderson disorder $U_j$ uniformly distributes in the range $[-W/2,W/2]$.
The localization of an eigenstate $|\varphi_j\rangle=\{ \phi_{j,x}\}_{x\in N_{\rm sites}}$ can be characterized by the inverse participation ratio (IPR)\cite{LiX17prb,Roy21prl,LiCA22prb}, defined as
\begin{equation}\label{}
  \operatorname{IPR}_j=\sum_{a=1}^{N_{\text {sites }}}\left|\phi^R_{j, a}\right|^4.
\end{equation}
The meaning of IPR can be understood in terms of a state uniformly distributed over $N$ sites, in which all non-vanishing elements of the wave function are $\phi_{j,a}=1/\sqrt{N}$. According to the above definition, we have ${\rm IPR}_j=1/N$ for the homogeneous state. Hence, ${\rm IPR}_j$ can measure the number of the distributed sites of an eigenstate ($1/{\rm IPR}_j$), which can be taken as a degree of its localization. Besides, since the wave function is normalized, the value is bounded by $0<{\rm IPR}_j<1$.

To see how the ${\rm IPR}_j$ measures the localization degree of an eigenstate explicitly, we first calculate the inverse participation ratio at various disorder strengths in the Hermitian case by setting $\gamma=0$. As shown in Fig. \ref{fig2}(c), ${\rm IPR}_j$s inside the two topological gaps take very small values (less than $10^{-3}$), which is expected as the topological edge states are delocalized over a large number of lattice sites. In addition, ${\rm IPR}_j$s of the edge states almost do not fluctuate, and have no noticeable change with increasing the disorder strengths. In contrast, the other states have much larger ${\rm IPR}_j$s, and exhibit significant fluctuations among different eigenstates and disorder strengths, which can be viewed as the signatures of the localization.

Now we turn on the non-Hermitian potential, and show the results of $\gamma=0.3$ in Fig. \ref{fig2}(d). It is found that the fluctuations of ${\rm IPR}_j$s remain very low inside the topological gaps, and the values only slightly vary as the disorder strength is increased, which are in stark contrast to those in the other regions. It implies the non-Hermitian edge states are still topologically protected, and they are robust against on-site disorder. However, when compared with the Hermitian counterpart, the values of ${\rm IPR}_j$s considered here in the topological gaps are significant increased, even in the absence of disorder. Hence, it implies possible localization of the edge states, which is related to a hybrid  skin effect happening at the boundary of the hyperbolic lattice, as we will show in the next section.

\begin{figure}[hbpt]
\centering
\includegraphics[width=8.6cm]{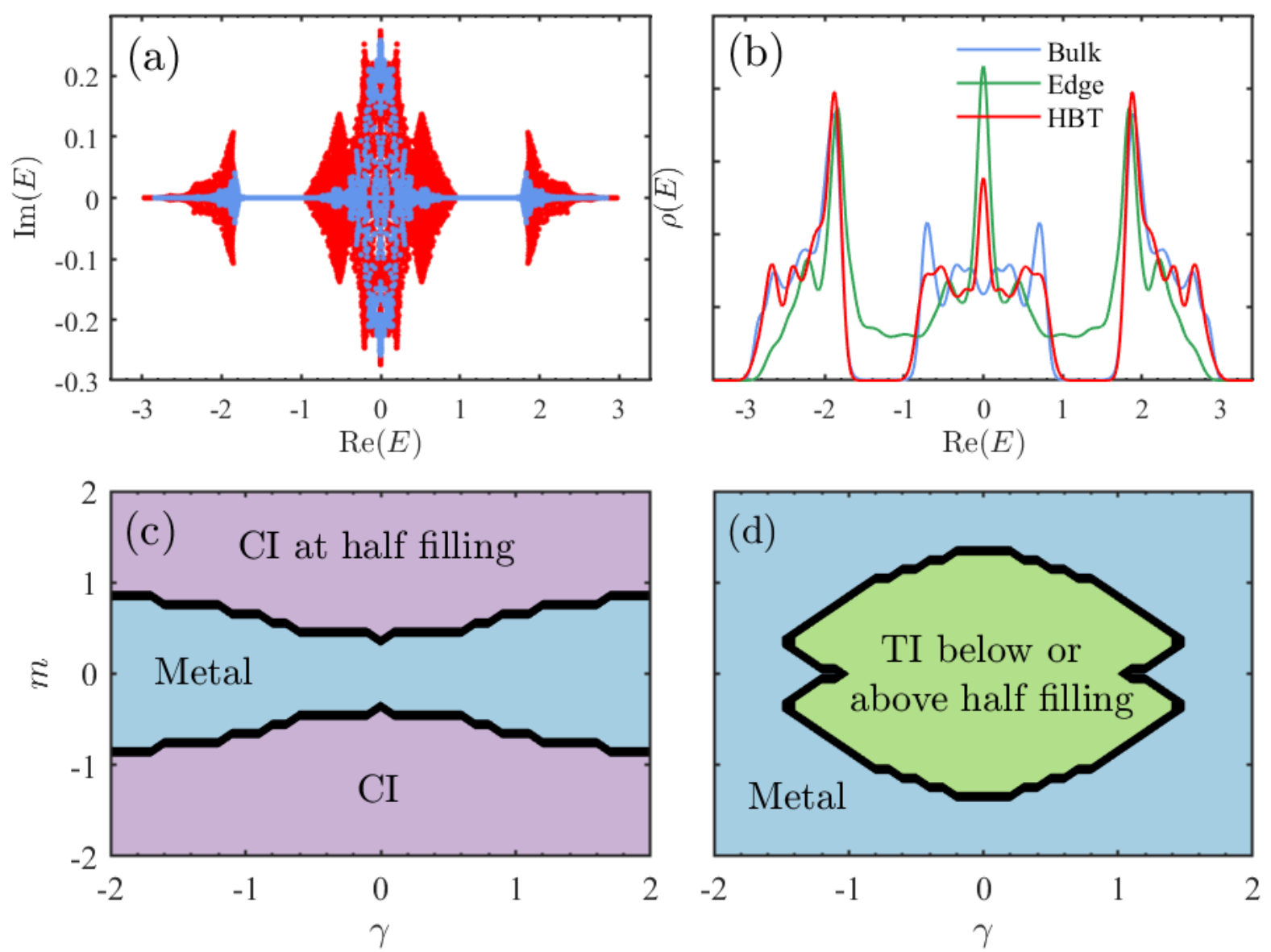}
\caption{(a) Energy spectrum and (b) the corresponding density of states of the Hamiltonian in Eq.(1)
on a hyperbolic $\{8,3\}$ lattice obtained by HBT (red dots or line) with periodic boundary condition. In (a) and (b), the same quantities calculated by directly diagonalizing the Hamiltonian of an open hyperbolic flake are also plotted for comparison.
The phase diagrams in the $(\gamma,m)$ plane: (c) at half filling; (d) below or above half-filling. CI (TI) represents conventional (topological) insulator. Here the parameters not specified are the same as those used in Fig.~\ref{fig1}.}\label{fig3}
\end{figure}

Recently, the Bloch band theory has been generalized to hyperbolic lattices\cite{doi:10.1126/sciadv.abe9170}. A hyperbolic Bravais lattice with periodic boundary condition can be constructed by applying Fuchsian translation on the unit cell. Instead of dealing with a large real-space Hamiltonian, the single-particle spectrum can be easily obtained by diagonalizing a Hamiltonian in the momentum space (this momentum space is at least four-dimensional, which is unlike that of a Euclidean lattice). For the hyperbolic $\{8,3\}$ lattice, we can choose a unit cell containing $16$ sites, thus have four independent generators to form a hyperbolic translation group (See Appendix \ref{AppendixA} for the details). In the four-dimensional momentum space $(k_1, k_2, k_3, k_4)$, the resulting hyperbolic Bloch Hamiltonian $\mathcal{H}_{\{8,3\}}(\boldsymbol{k})$ is a $16\times16$ matrix (see Appendix \ref{AppendixB} for the explicit form). Since the hyperbolic band theory directly solves the bulk system, the phase diagram in the $(\gamma,m)$ parameter space can be determined more precisely without the mixing of the edge states. Figure \ref{fig3}(a) shows the boundaries between the metal and conventional insulator (CI) at half filling. A CI can only be induced by a large enough staggered potential $m$, and the critical value $m_c$ increases with the strength of the non-Hermitian potential $\gamma$.

While the topological state is absent at half filling, it appears in symmetric locations below and above the $\mathrm{Re}(E)=0$, as shown in Fig. \ref{fig3}(b). For a  small $\gamma$, the topological property will be broken by a finite $m$. The critical value $m_c$ decreases with increasing the absolute value of $\gamma$. As $\gamma$ becomes large enough, the topological phase is completely suppressed, and the system keeps metallic at any value of $m$. Interestingly, near the left or right boundaries of the topological region, the topological phase can be induced by turning on $m$. A qualitative understanding is that the effect of the topological hopping term is manifested due to the annihilation between the real and imaginary staggered potentials.

\begin{figure}[htbp]
\centering
\includegraphics[width=8.6cm]{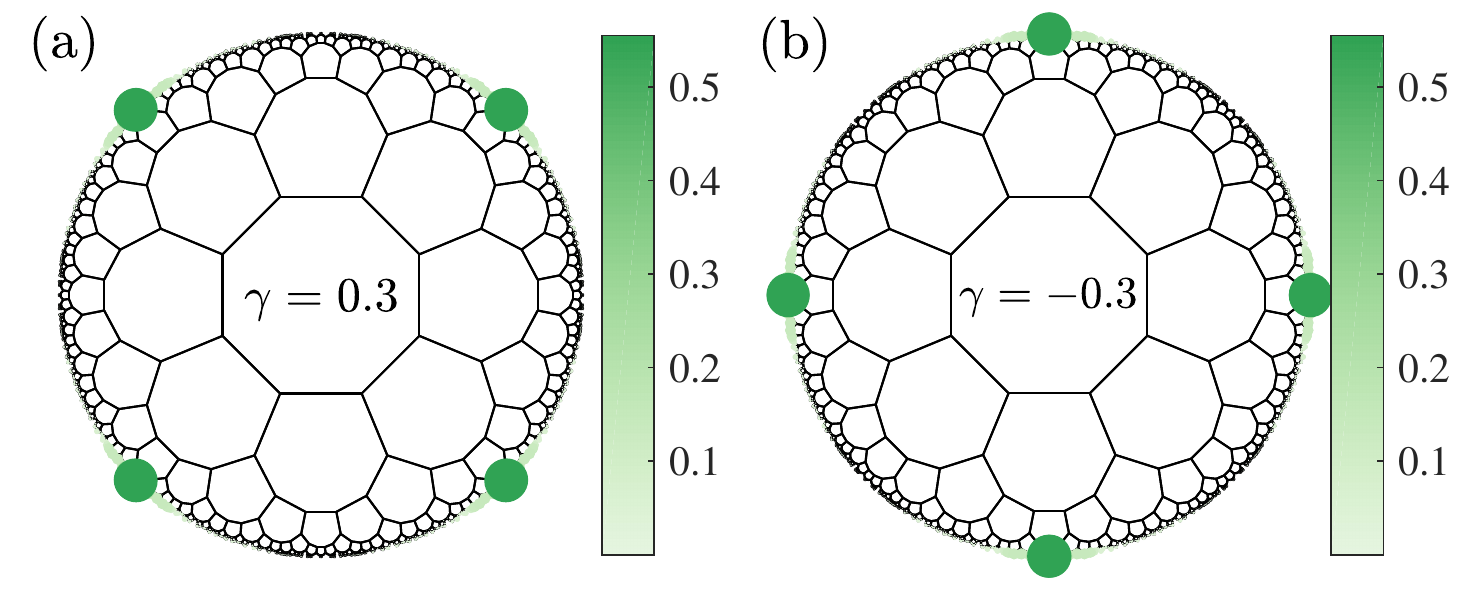}
\caption{Demonstration of the skin-topological corner modes in terms of the sum of the distributions of the edge states in the energy range $1-1.4 ~t_1$ for (a) $\gamma=0.3$ and (b) $\gamma=-0.3$. The other parameters are the same as those in Fig.~\ref{fig1}.}\label{fig4}
\end{figure}

\begin{figure*}[htbp]
\centering
\includegraphics[width=16cm]{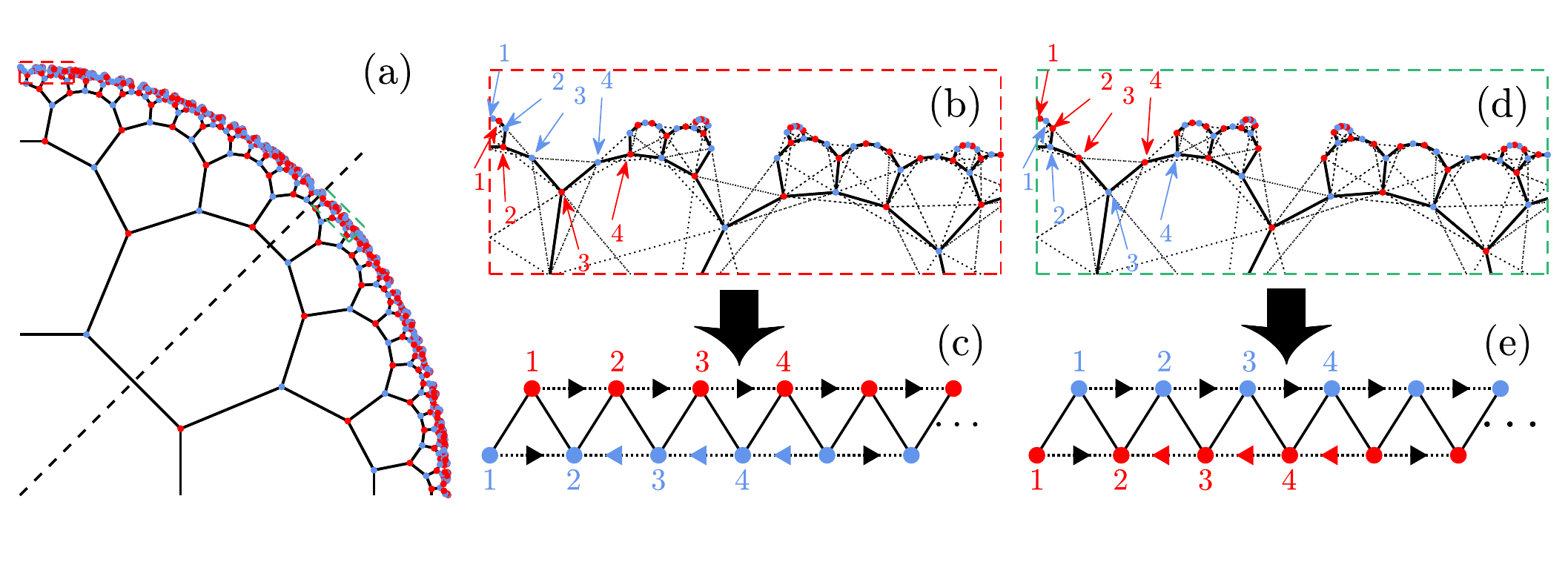}
\caption{(a) Illustration of the inequivalent one-fourth of the studied open hyperbolic lattice. Zoomed details of the edge geometries in the dotted line boxes located at: (b) the top (red); (d) the diagonal (green). (c) and (e) show the outer most sites arranged in zigzag shapes for the boundary fragments (b) and (d), respectively. The arrows in (c) and (e) denote the directions along which the next nearest-neighbor hopping phase $\phi$ is positive.}\label{fig5}
\end{figure*}

\section{Hybrid higher-order skin-topological effect}
\label{sec3}
In this section, we investigate the effect of the non-Hermitian potential on the edge states in the topological gap. The on-site gain and loss drive the boundary states to localize at four corners of the disk geometry, realizing hybrid higher-order skin-topological effect.
As shown in Fig.~\ref{fig4}, the corner states are located at the diagonal positions for $\gamma>0$, and axial positions for $\gamma<0$.
The geometry considered here has an eight-fold rotation symmetry, and can be divided into eight equal circular sectors with central angle $\pi/4$. The corners can be regarded as the boundary sites where two adjacent pieces meet. After the staggered potential is included, the rotation symmetry is reduced to be four-fold, which accounts for the existence of localized states at four symmetric corners.

We then proceed to the underlying mechanism causing the localization of corner states.
The phase $\phi$ in the NNN hopping results from an alternating magnetic flux piercing the hyperbolic plane perpendicularly. Similar to the celebrated Haldane model on honeycomb lattice, the total magnetic flux in each octagon is completely canceled, and sums up to zero. However, since the boundary is composed of open-sided octagon, there exist net magnetic fluxes nearby. In the following, we will show the interplay between the non-vanishing boundary flux and the non-Hermitian physics generate a novel skin effect, which results in the appearance of the corner modes.

For simplicity, we ignore the connections between the inner sites and the boundary ones, and model the edge state with a separate zigzag chain [see Fig.~\ref{fig5}(c)]. The phase $\phi$ of the NNN hopping amplitude is positive when the electron moves in the arrow direction. If the positive directions for the two sublattices are the same and the numbers of the NNN bonds in the upper and lower chains are equal, the phases picked up by an electron circling the zigzag area through all NNN bonds are exactly canceled out, suggesting there is no net magnetic flux piercing the region. This situation occurs at a zigzag edge of the Euclidean honeycomb lattice. However, in the current study, we have an irregular boundary, which results from constructing the disk geometry using complete unit cells with the centers in a fixed radius. As shown in Fig.~\ref{fig5}, the positive directions are not uniform any more [to be clear, the positive direction pointing to left is denoted by red (blue) color in the upper (lower) sublattice]. The net magnetic flux is proportional to the difference in the numbers of the red arrows and the blue ones. Specifically, we focus on one-fourth of the circle geometry due to the four-fold symmetry, as demonstrated in Fig.~\ref{fig5}(a). Without the staggered potential, the upper half sector is exactly the same as the lower half one, and they are related to each other by a $\pi/4$ rotation. After $\gamma$ is included, the two sectors become inequivalent, and the difference is that the signs of the imaginary staggered potentials on the two sublattices are interchanged in the lower sector. The skin effect induced by a $\gamma>0$ non-Hermitian term manifests itself in the upper sector by making all modes localize at the right end [Figs.~\ref{fig5}(b) and \ref{fig5}(c)]. In contrast, due to the sign reversing of the non-Hermitian staggered potential in the lower sector, the localization from the skin effect occur at the left end [Figs.~\ref{fig5}(d) and \ref{fig5}(e)]. Hence, the two opposite localization tendencies meet at the diagonal position, generating a skin-topological corner state there. For the case of $\gamma<0$, the localization of the skin mode changes to the opposite end of each $1/8$ sector, and the corner states appear in the axial positions.

\begin{figure}[hbpt]
\centering
\includegraphics[width=8.6cm]{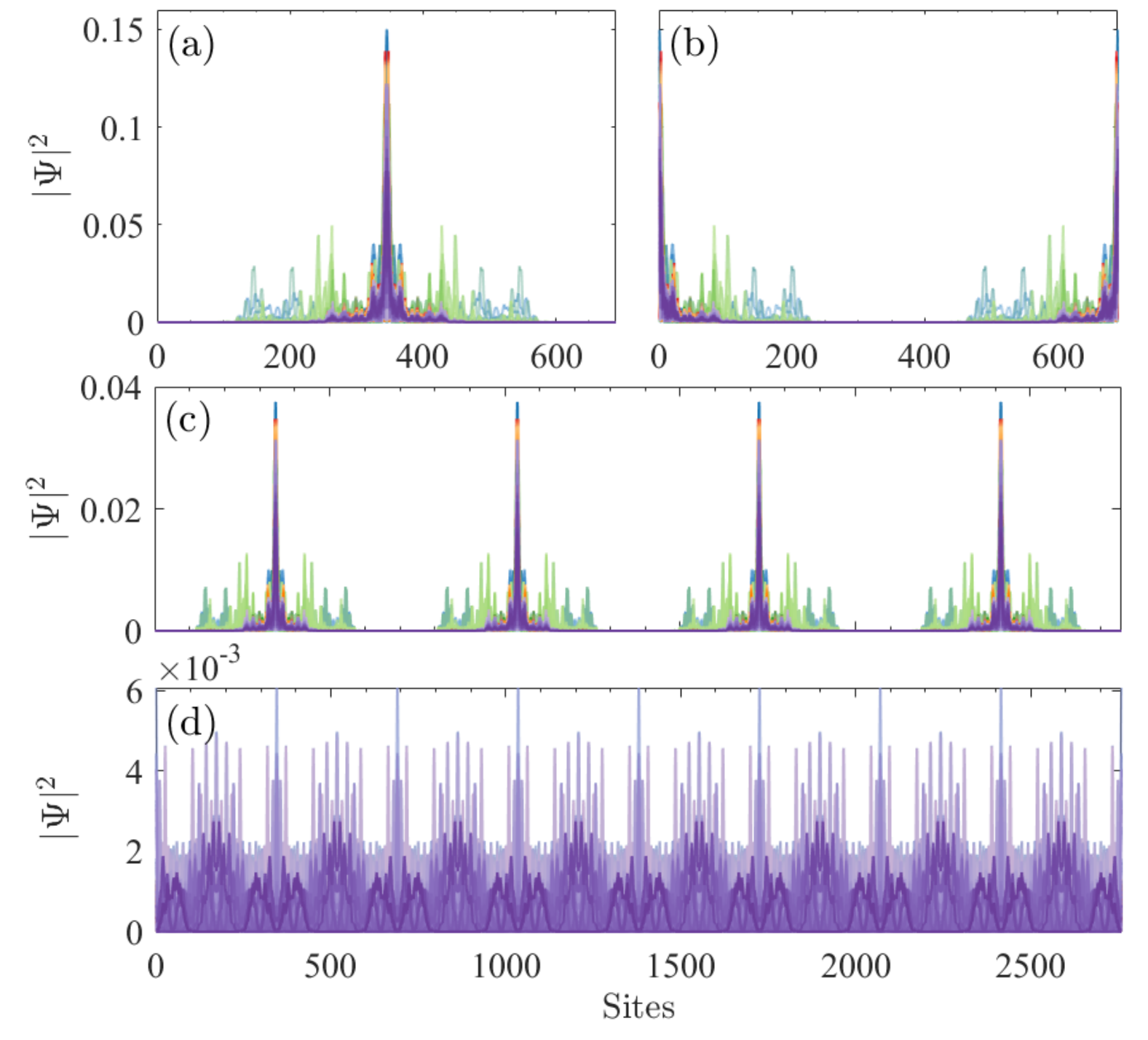}
\caption{The distributions of the eigenstates of the one-dimensional zigzag chain model mapped from the boundary of the inequivalent one-fourth geometry in Fig.~\ref{fig5}(a) at (a) $\gamma=0.8$ and (b) $\gamma=-0.8$. (c) and (d) are the distributions on a closed one-dimensional chain corresponding to the boundary of the whole hyperbolic disk with $\gamma=0.8$ and $\gamma=0$, respectively. The eigenstates are distinguished by different colors in (a)-(d).}\label{fig6}
\end{figure}

The above skin effect can be explicitly demonstrated by examining the distributions of the eigenstates of the effective zigzag-chain Hamiltonian. We first diagonalize the non-Hermitian model mapped from the boundary of the geometry in Fig.~\ref{fig5}(a), and find the tendency for all eigenstaes to distribute near the middle ($\gamma>0$) or ends ($\gamma<0$) of the one-dimensional(1D) zigzag chain [Figs.~\ref{fig6}(a) and \ref{fig6}(b)]. Then we proceed to consider a closed zigzag chain, which corresponds to the whole boundary of a hyperbolic disk. As shown in Fig.~\ref{fig6}(c), there are four skin-topological corner modes, and their locations are in good consistence with those obtained by the calculations on the entire lattice. The skin corner modes are induced by the imaginary staggered potential. To verify this, we plot the distributions of the eigenstates in Fig.~\ref{fig6}(d), and indeed the corner states do not appear any more without the non-Hermitian potential.

\begin{figure}[hbpt]
\centering
\includegraphics[width=8.6cm]{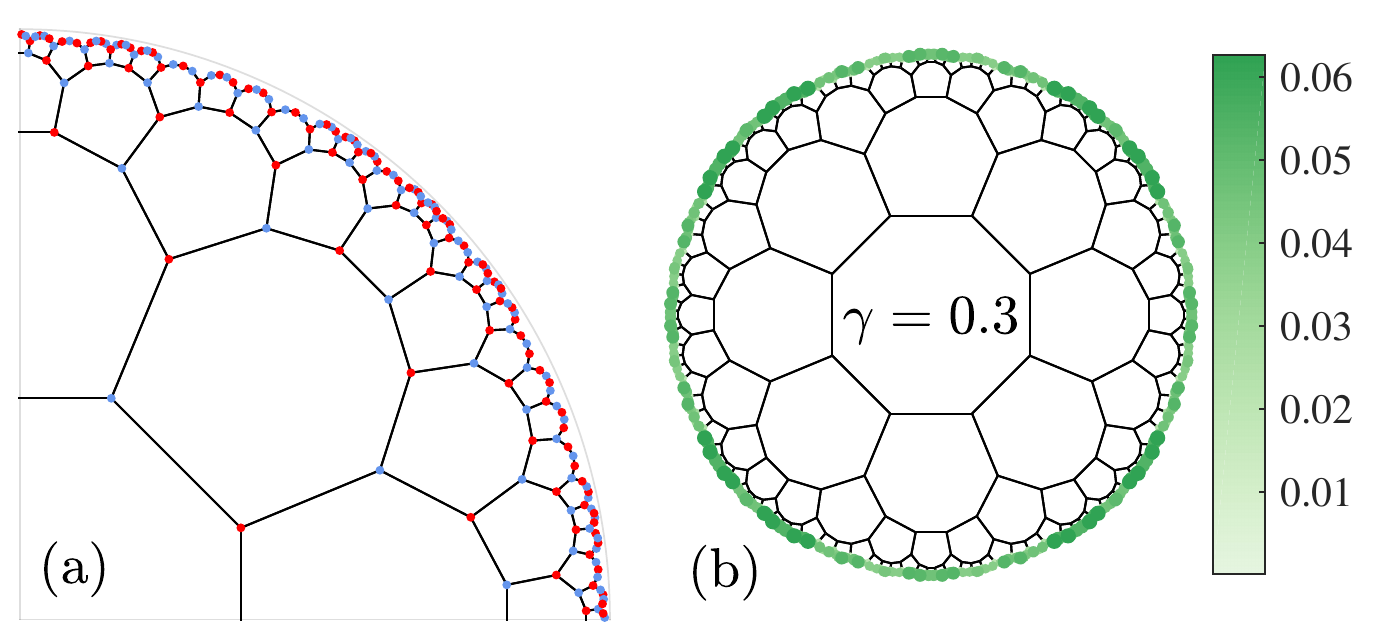}
\caption{(a) A smoother boundary created by cutting the hyperbolic lattice with a fixed-radius circle. (b) The distributions of the edge states in
the topological energy gap at $\gamma=0.3$, and no skin-topological corner modes are visible. The other parameters are the same as those in Fig.~\ref{fig1}.}\label{fig7}
\end{figure}

It is worth noting that the geometry of the boundary is critical for the appearance of the skin-topological corner modes. If the open hyperbolic flake is constructed by keeping the lattice sites located within a fixed-radius circle, the effective 1D zigzag chain is not threaded by a net magnetic flux. Consequently, the non-Hermitian potential will not lead to the skin effect, and no skin corner states are generated (see Fig.~\ref{fig7}).

\section{Non-Hermitian Haldane model on hyperbolic \{12,3\} lattice}
\label{sec4}
\begin{figure}[htbp]
\centering
\includegraphics[width=8.7cm]{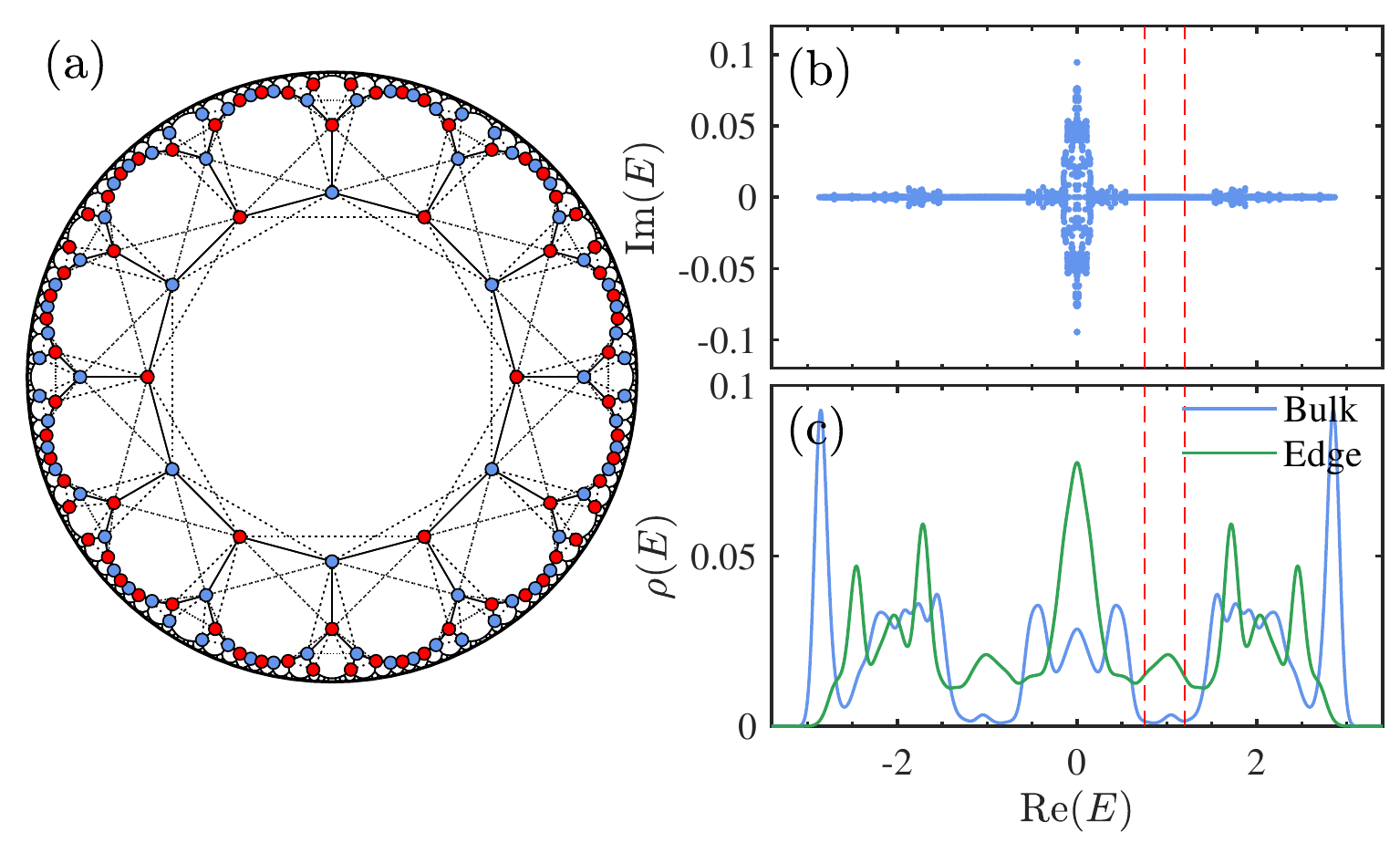}
\caption{(a) Schematic illustration of the Haldane model on a hyperbolic $\{12, 3\}$ lattice. (b) Energy spectrum of the non-Hermitian Haldane Hamiltonian on an open hyperbolic $\{12, 3\}$ lattice. (c) The corresponding density of states of bulk (blue) and edge (green) states. The red dotted lines in (b) and (c) estimate the topological gap above the $\mathrm{Re}(E)=0$ (the lower one is symmetric). Here the parameters in (b) and (c) are $t_1=1$, $t_2=0.4$, $m=0$ and $\gamma=0.1$. }\label{fig8}
\end{figure}

The hyperbolic space has infinitely number of regular tilings, each of which can be labeled by the so-called Schl\"afli symbol $\{p,q\}$ with $p$ the number of sides of the polygons and $q$ the coordination number at each vertex. Hence, we proceed further to investigate more honeycomb-like hyperbolic lattices with $p>8$ while fixing $q=3$ to see the evolution of the physical properties of the non-Hermitian Haldane model with a polygon geometry [see Fig.~\ref{fig8}(a)]. The main physics, such as the topological nontrivial phase and hybrid higher-order skin-topological effect, are similar to the hyperbolic $\{8,3\}$ presented above.  For instance, the NNN spin-orbit coupling hoppings also lead to two topologically nontrivial band gaps here, which distribute symmetrically with respect to $\mathrm{Re}(E)=0$ [Figs.~\ref{fig8}(b) and \ref{fig8}(c)]. However, a larger $t_2$ is needed to generate a gap with the same size in the hyperbolic $\{12,3\}$ lattice. When the non-Hermitian term $\gamma$ is introduced, the energy spectrum becomes complex, and the eigenvalues come in pairs due to the pseudo-Hermicity of the Hamiltonian. As shown in Fig.~\ref{fig8}(c), the boundary states remain within the topological energy gap in the presence of the non-Hermitian potential, implying the topological property persists under non-Hermitian effect. We find that the real-space Chern number gets quantized in the topological gaps. Besides, these non-Hermitian edge states corresponding to nontrivial phases show robustness against on-site disorder, which suggests they are topologically protected.

\begin{figure}[htbp]
\centering
\includegraphics[width=8.7cm]{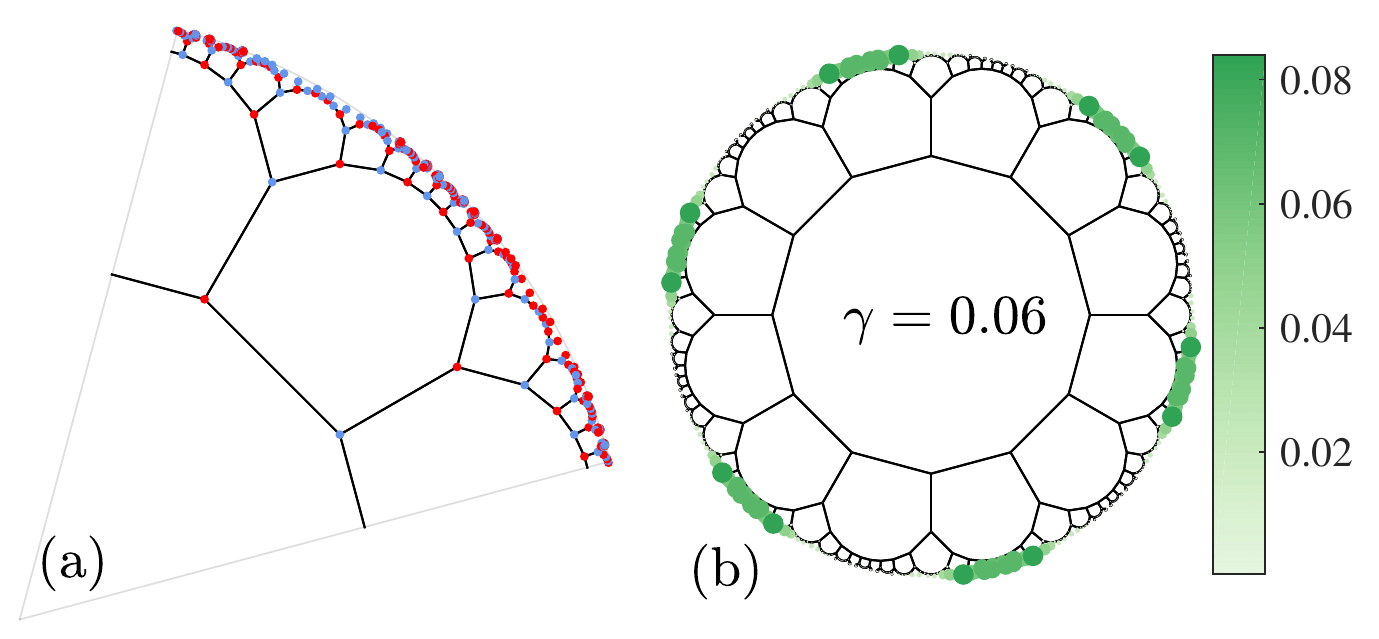}
\caption{(a) The inequivalent one-sixth of the open hyperbolic $\{12,3\}$ lattice. (b) The skin-topological modes appearing symmetrically at six corners of an open hyperbolic $\{12, 3\}$ lattice.}\label{fig9}
\end{figure}

Then we consider a disk of hyperbolic $\{12,3\}$ lattice with an open boundary while keeping the unit cells whose centers are within a fixed radius. The positive directions of the imaginary NNN hoppings are imbalanced on the resulting boundary, near which there is a net magnetic flux threading through. Consequently, the skin-topological modes appear at six corners of the disk geometry. The six corners arise due to the $6$-fold rotational symmetry in the hyperbolic $\{12,3\}$ flake. The   underlying mechanism is similar as in the hyperbolic $\{8,3\}$ case discussed previously. Let us focus on one-sixth of the geometry [see Fig.~\ref{fig9}(a)]. Actually, this sector can be further divided into two equal parts, whose Hamiltonians are almost the same except the signs of the staggered potentials. The difference leads to the directions of the skin effects are opposite, thus generating one corner mode in each one-sixth sector [see Fig.~\ref{fig9}(b)]. Compared to the hyperbolic $\{8,3\}$ case, the corner modes here are more extended, which is due to the smaller net magnetic flux near the $\{12,3\}$ boundary.

\section{Conclusions and Discussion}
\label{sec5}

We have investigated the topological properties of non-Hermitian Haldane model on hyperbolic lattices of two representative $\{8, 3\}$ and $\{12, 3\}$ types. We prove the existence of non-trivial topological gap in the complex energy spectrum, charaterized by a quantized non-Hermitian Chern number. Correspondingly, the edge states appear at boundaries of the hyperbolic lattices in the topological non-trivial phase. Remarkably, these edge states are driven to localize at corners of the boundary by non-Hermitian effect, forming the hybrid higher-order skin-topological effect.  The mechanism accounting for such a hybrid higher-order skin-topological effect is revealed by mapping the specific boundary to an effective one-dimensional zigzag chain model with a net magnetic flux. We expect the exotic non-Hermitian topology on hyperbolic lattices can be experimentally demonstrated, for instance, in electric circuits\cite{RN55,RN56,RN57,Zhang23nc}.

While we focus on hyperbolic lattices of $\{8,3\}$ and $\{12,3\}$ types here, the main results can be generalized to other types of different $\{p,q\}$ with similar setup.  We note the number of degrees of freedom at the boundary of hyperbolic lattices is very large, different from regular lattices, such that the ratio of boundary modes over bulk modes is extremely high. Therefore, the hybrid higher-order skin-topological effect may provide an effective method to manipulate most degrees of freedom in  hyperbolic lattices.

\appendix

\renewcommand{\thefigure}{A\arabic{figure}}
\setcounter{figure}{0}
\section{Hyperbolic Geometry}\label{AppendixA}
In this appendix, we present the general description of the hyperbollic geometry. The  hyperbolic plane is a two-dimensional  space  with  negative  curvature.  An  infinitely periodic  lattice  on  this  plane  can  be  mapped  onto  a  unit  disk $D=\{z\in \mathbb{C}, |z|<1\}$,  known  as  the  Poincar\'{e}-disk  representation\cite{BALAZS1986109,PhysRevB.105.125118}.  The  distance  metric  on  the  Poincar\'{e}  disk  is
\begin{equation}\label{}
  {\rm d}s^2=(2\kappa)^2\frac{ {\rm d}x^2+ {\rm d}y^2}{(1-|z|^2)^2},
\end{equation}
where  $\kappa$  represents  the  radius  of  curvature.  From this  metric,  it  is clear  that  the  hyperbolic  distance  between two  points close to each other  on  the  edge  of  the  unit  disk  in  the  Poincar\'{e}  representation  is  infinite. Explicitly,  it  can  be  inferred  that  the  hyperbolic  distance  between  any  two  points,  $z_1$  and  $z_2$,  on  the  unit  disk  is
\begin{equation}\label{distance}
  d(z_1,z_2)=\kappa{\rm arcosh}\left(1+\frac{2|z_1-z_2|^2}{(1-|z_1|^2)(1-|z_2^2)}\right).
\end{equation}
For  the hyperbollic $\{p,q\}$  lattice  on  the  hyperbolic  plane,  the  interior  angle  of  each  regular  p-gon  is  $2\pi/q$.  If  a  regular  $p$-gon  is  placed  with  its  center  at  the  origin  of  the  Poincar\'{e}  disk,  then  the  distance  from  a  vertex  of  the  polygon  to  the  center  of  the  disk  is  \begin{equation}\label{}
  r_0=\sqrt{\frac{\cos(\pi/p+\pi/q)}{\cos(\pi/p-\pi/q)}}.
\end{equation}
The  coordinates  of  these  vertices  can  be  expressed  as
$z_j=r_0e^{{\rm i}(2\pi j/p+\delta)}$, where$ \ j=1,\ldots,p$ with $\delta$ an arbitrary phase factor.

\begin{figure}
  \centering
  \includegraphics[width=8.6cm]{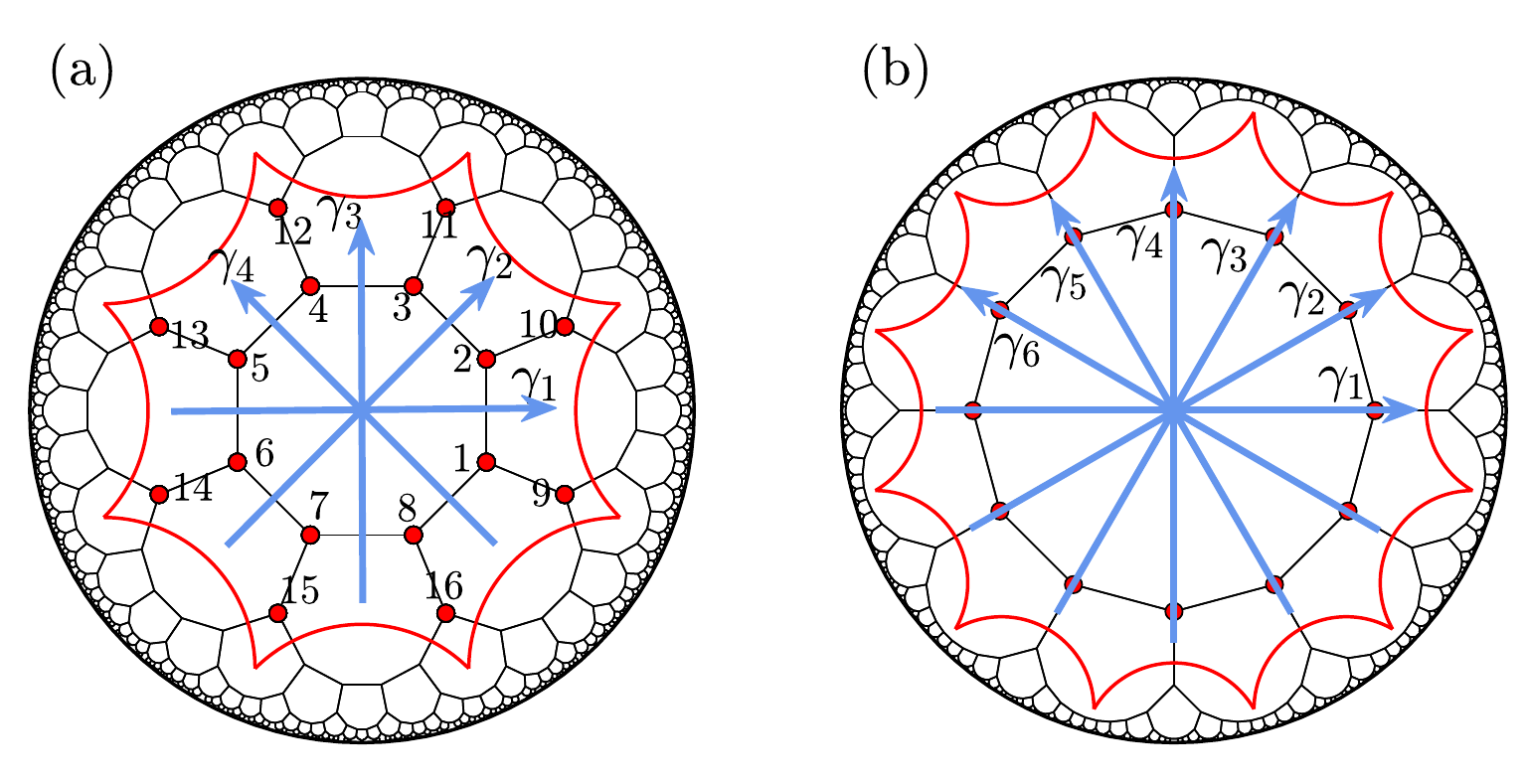}\\
  \caption{ Sketch of hyperbolic $\{8,3\}$ [panel (a)] and $\{12,3\}$ [panel (b)] lattices with marked unit cell (red dots) and fundamental domain of the Bravais lattice (red polygon), respectively. $\gamma_j$ in (a) and (b) are the generators of their lattice.}\label{figA1}
\end{figure}

For  the  hyperbollic $\{8,3\}$  lattice,  its  unit  cell  consists  of  a  regular  octagon  indicated  by  the  red  lines  in  Fig.~\ref{figA1}(a),  containing  16 sites.  Therefore,  the  $\{8,3\}$  lattice  can  be  tessellated  by  $\{8,8\}$  lattice. Similarly,  for  $\{12,3\}$  lattice, one  unit  cell  consists  of  a  regular  dodecagon  containing  12  sites such that the  $\{12,3\}$  lattice  can  be  tessellated  by  $\{12,12\}$  lattice. As  long  as  the  coordinates  of  the  sites  contained  in  an  initial  unit  cell  are  given,  the  coordinates  of  all  other sites  on  the  entire  lattice  can  be  obtained  by  translation  operations  of  the  sites  within  the  unit  cell.  That  is,  any  site  can  be  uniquely  represented  as
\begin{equation}\label{}
  z_i=\gamma z^{(a)},
\end{equation}
where  $a$  is  the  index  of  the  site  within  the  unit  cell (for $\{8,3\}$ latice, $a=1,\cdots,16$),  and  $\gamma$  is  the  sequence  of  basic  translation  operations  that  make  transformations  between  different  unit  cells. The  translational  operator  in  the  horizontal  direction  of  the  hyperbolic  plane  can  be  represented  by  the  matrix
\begin{equation}\label{}
  T(\tau)=\left(
            \begin{array}{cc}
              \cosh(\tau/(2\kappa)) & \sinh(\tau/(2\kappa)) \\
              \sinh(\tau/(2\kappa)) & \cosh(\tau/(2\kappa)) \\
            \end{array}
          \right),
\end{equation}
with  $\tau$  representing  the  hyperbolic  distance  translated  horizontally  by  $T(\tau)$.  For  instance,  by  applying  the  translation  operator  $T(\tau)$  to  the  origin  $0$,  we  can  utilize  Eq.~\ref{distance}  to  calculate  the  distance  $d(z,0)$  between  a  point  $z$  and  the  origin, allowing  us  to  discern  the  distance between  the  origin $0$  and  $T(\tau)0$ is   $\tau$.  This  is  critical  for  understanding  the  parameter $\tau$ in the horizontal  translation operation $T(\tau)$.

The  unit  cell  on  the  $\{8,  3\}$  lattice  contains  four  fundamental  translation  symmetry  operations,  also  known  as  generators,  namely  $\gamma_m$$ (m=1,2,3,4)$, as  shown  in  Fig.~\ref{figA1}.  Among  them,  the $\gamma_1$  stands  for  the  operation  that  horizontally  translates  a  unit  cell  to  its  nearest  neighbor  on  the  right.
Assuming that the hyperbolic distance from the center of the Bravais lattice to the center of a side is $a$, then the translation distance of $\gamma_1$ is $2a$, therefore
\begin{equation}\label{}
  \gamma_1=T(2a)=\left(
            \begin{array}{cc}
              \cosh(2a/(2\kappa)) & \sinh(2a/(2\kappa)) \\
              \sinh(2a/(2\kappa)) & \cosh(2a/(2\kappa)) \\
            \end{array}
          \right),
\end{equation}
and the length of $a$ satisfies
\begin{equation}\label{}
  \cos\left(\frac\pi p\right)=\frac{\tanh[2{\rm arctanh}(a)]}{\tanh[2{\rm arctanh}(r_0)]}.
\end{equation}
As long as $\gamma_1$ is known, other generators can be directly obtained through a rotation matrix
\begin{equation}\label{}
  R(\phi)=\left(
            \begin{array}{cc}
              e^{{\rm i}\phi/2} & 0 \\
              0 & e^{-{\rm i}\phi/2} \\
            \end{array}
          \right).
\end{equation}
Henceforth,
\begin{equation}\label{}
  \gamma_{m=1,2,3,4}=R((m-1)\alpha)\gamma_1R(-(m-1)\alpha)
\end{equation}
with $\alpha=\frac{2\pi}{p_B}$.

All  the  translation  symmetry  operations  of  the  unit cell  form  a  Fuchsian  group  $\Gamma_g$.  Each  element  in  $\Gamma_g$  is  a  sequence  of  operators  consisting  of  generators  and  their  inverses.  For  convenience,  the  inverse  of  a  generator  is  also  labeled  as  a  generator,  denoted  by  $\gamma_{p_B/2+j}=(\gamma_m)^{-1}$. Then the generator contains $\gamma_m$ with $m=1,\ldots,p_B$. Each $\gamma$ is a word of a specific length composed of generators, for example a word of length $n$ can be written as
\begin{equation}\label{}
  \gamma=\gamma_{m_1}\cdots\gamma_{m_n}
\end{equation}
with $m_i=\{1,\ldots,p_B\}$. What matters for $\gamma$ is only the sequence of generators such that a $\gamma$ can be labeled with a sequence of numbers $\vec m=(m_1,\ldots,m_n)$. In general, any site in the lattice can be uniquely written as $(a,\vec m)$.

\section{Hyperbolic Bloch Hamiltonian}\label{AppendixB}
In this appendix, we present the explicit form of Boloch Hamiltonian for hyperbollic lattice. The hyperbolic Haldane Bloch Hamiltonian ${\mathcal H}_{\{8,3\}}({\bm k})$ is a $16\time 16$ matrix, which can be expressed as
\begin{equation}\label{}
  {\mathcal H}_{\{8,3\}}({\bm k})=H^1+({H^1})^\dagger+H_S.
\end{equation}
By  defining  $S=m+i\gamma$  and  $f=e^{i\phi}$,  the  specific  forms  of  matrices  $H_S$  and  $H^1$  are  given  by
\begin{widetext}
\begin{equation}\label{}
  H_S={\rm diag}(S[-1,1,-1,1,-1,1,-1,1,1,-1,1,-1,1,-1,1,-1])
\end{equation}
\begin{align}\label{}
\nonumber
&H^1_{2, 1}= t_1;
H^1_{3, 2}= t_1;
H^1_{4, 3}=t_1;
H^1_{5, 4}=t_1;
H^1_{6, 5}= t_1;
H^1_{7, 6}= t_1;
H^1_{8, 7}= t_1;
H^1_{1, 8}= t_1;
\\ \nonumber
&H^1_{9, 1}= t_1;
H^1_{2, 10}= t_1;
H^1_{11, 3}= t_1;
H^1_{4, 12}= t_1;
H^1_{13, 5}= t_1;
H^1_{6, 14}= t_1;
H^1_{15, 7}= t_1;
H^1_{8, 16}= t_1;
\\ \nonumber
&H^1_{9, 14}= t_1e^{i k_1};
H^1_{10, 15}= t_1e^{i k_2};
H^1_{11, 16}= t_1e^{i k_3};
H^1_{12, 9}= t_1e^{i k_4}; \\ \nonumber
&H^1_{13, 10}= t_1e^{-i k_1};
H^1_{14, 11}= t_1e^{-i k_2};
H^1_{15, 12}= t_1e^{-i k_3};
H^1_{16, 13}= t_1e^{-i k_4};
\\ \nonumber
&H^1_{3, 1}= t_2f;
H^1_{4, 2}= t_2f;
H^1_{5, 3}= t_2f;
H^1_{6, 4}= t_2f;
H^1_{7, 5}= t_2f;
H^1_{8, 6}= t_2f;
H^1_{1, 7}= t_2f;
H^1_{2, 8}= t_2f;
\\ \nonumber
&H^1_{1, 10}= t_2f;
H^1_{2, 11}= t_2f;
H^1_{3, 12}= t_2f;
H^1_{4, 13}= t_2f;
H^1_{5, 14}= t_2f;
H^1_{6, 15}= t_2f;
H^1_{7, 16}= t_2f;
H^1_{8, 9}= t_2f;
\\ \nonumber
&H^1_{9, 2}= t_2f;
H^1_{10, 3}= t_2f;
H^1_{11, 4}= t_2f;
H^1_{12, 5}= t_2f;
H^1_{13, 6}= t_2f;
H^1_{14, 7}= t_2f;
H^1_{15, 8}= t_2f;
H^1_{16, 1}= t_2f;
\\ \nonumber
&H^1_{14, 1}= t_2fe^{-ik_1};
H^1_{15, 2}= t_2fe^{-ik_2};
H^1_{16, 3}= t_2fe^{-ik_3};
H^1_{9, 4}= t_2fe^{-ik_4};\\ \nonumber
&H^1_{10, 5}= t_2fe^{ik_1};
H^1_{11, 6}= t_2fe^{ik_2};
H^1_{12, 7}= t_2fe^{ik_3};
H^1_{13, 8}= t_2fe^{ik_4};
\\ \nonumber
&H^1_{1, 12}= t_2fe^{-ik_4};
H^1_{2, 13}= t_2fe^{ik_1};
H^1_{3, 14}= t_2fe^{ik_2};
H^1_{4, 15}= t_2fe^{ik_3};\\ \nonumber
&H^1_{5, 16}= t_2fe^{ik_4};
H^1_{6, 9}=t_2fe^{-ik_1};
H^1_{7, 10}= t_2fe^{-ik_2};
H^1_{8, 11}= t_2fe^{-ik_3};
\\ \nonumber
&H^1_{9, 11}= t_2fe^{-ik_2 + ik_1};
H^1_{10, 12}= t_2fe^{-ik_3 + ik_2};
H^1_{11, 13}= t_2fe^{-ik_4 + ik_3};
H^1_{12, 14}= t_2fe^{ik_1 + ik_4};\\
&H^1_{13, 15}= t_2fe^{ik_2 - ik_1};
H^1_{14, 16}= t_2fe^{ik_3 - ik_2};
H^1_{15, 9}= t_2fe^{ik_4 - ik_3};
H^1_{16, 10}= t_2fe^{-ik_1 - ik_4}.
\end{align}
\end{widetext}
For  the  sake  of  conciseness,  only  the  non-zero  matrix  elements  of  $H^1$  are  presented  here, and the  numbering  scheme  of  the  sites  within  the  corresponding  unit  cell  is  illustrated  in  Fig.  A1(a).  The  rest  of  the  matrix  elements  are  zero.

\section*{Acknowledgments}
J.S. and H.G. acknowledges support from the NSFC grant Nos.~11774019 and 12074022. C.A.L. acknowledges financial support from Würzburg University.
S.F. is supported by the National Key Research and Development Program of China under Grant No. 2021YFA1401803, and NSFC under Grant Nos. 11974051 and 12274036.

%
%
%
%
%
%
%
\bibliography{hyperbolicref}

\end{document}